# Multichannel Nonlinear Equalization in Coherent WDM Systems based on Bi-directional Recurrent Neural Networks

Stavros Deligiannidis, Kyle R. H. Bottrill, Kostas Sozos, Charis Mesaritakis, Periklis Petropoulos, *fellow member Optica* and Adonis Bogris, *senior member Optica*

*Abstract*— Kerr nonlinearity in the form of self- and cross-phase modulation imposes a fundamental limitation to the capacity of wavelength division multiplexed (WDM) optical communication systems. Digital back-propagation (DBP) is a widely adopted technique for the mitigation of impairments induced by Kerr nonlinearity. However, multi-channel DBP is too complex to be implemented commercially in WDM systems. Recurrent neural networks (RNNs) have been recently exploited for nonlinear signal processing in the context of optical communications. In this work, we propose multi-channel equalization through a bidirectional vanilla recurrent neural network (bi-VRNN) in order to improve the performance of the single-channel bi-VRNN algorithm in the transmission of WDM polarization multiplexed signals. We compare the proposed digital algorithm to full-field DBP and to the single channel bi-RNN in order to reveal its merits with respect to both performance and complexity. We finally provide experimental verification in QPSK transmission, showcasing over 2.5 dB optical signal-to-noise ratio (OSNR) gain and a significant reduction in bit-error-rate (BER) by an order of magnitude compared to single channel equalization. Moreover, a complexity reduction of up to 43% was achieved compared to the single-channel RNN, and up to 550% compared to the single channel DBP in the case of 24-span long haul transmission.
© 2023 The Authors

*Index Terms*—Optical Fiber Communication, Nonlinear Signal Processing, Recurrent Neural Networks, Cross-Phase Modulation, Coherent Communication

## I. INTRODUCTION

COMMUNICATION rates exploiting optical fiber capacity has shown a remarkable exponential growth for more than 30 years, with the latest deployed systems operating even below a decibel from Shannon's limit [1], [2]. This pace is attributed to the evolution of optoelectronic devices that led to the adoption of coherent technology operating at very high bandwidth per channel and the advanced digital signal processing (DSP), in the form of probabilistic shaping [3], forward error correction codes [4] and digital compensation of the transmission impairments [5]. Periodically amplified, wavelength division multiplexed (WDM) systems covering medium to long distances are characterized by distributed non-linear effects, the most dominant of which arising from the intensity dependent refractive index. The well-known Kerr effect manifests itself as self-phase modulation (SPM) [6], cross-phase modulation (XPM) [7] or four wave mixing (FWM) [8] and is responsible for intra-channel distortions and inter-channel crosstalk. Intra-channel nonlinearity can be ideally compensated for either by digital back propagation (DPB) using the split-step Fourier method [9] or through the nonlinear Fourier transform [10].

Assuming ideal compensation of all intra-channel effects through DBP, XPM appears to be the principal source of impairments that fundamentally limits the information capacity of an optical communication system. The various techniques for nonlinear equalization treat XPM as a time-varying intersymbol interference (ISI) process and concentrate mainly in the zeroth-order XPM contribution [11], [12], known as phase and polarization-rotation noise. In this way, by tracking the temporal changes of the ISI, linear or turbo equalizers [11], [13] can partially mitigate its effects. On the other hand, higher order XPM contributions cannot be efficiently equalized unless very complex multi-channel full-field DBP [14]–[16] or Volterra equalizers [17] are employed. However, these multi-channel equalizers increase significantly the power consumption of the DSP and thus, they are quite inapplicable. The coupled-channel DBP [18] aims to reduce the complexity of full-field DBP by separately representing the WDM channels and explicitly accounting for their interaction during propagation. However, even with this simplification and the subsequent performance sacrifice, the complexity issue remains. Many works have concentrated on complexity savings by reducing the computational steps per span in DBP [19], [20], but this may be the case only for single-channel DBP, while the multi-channel implementations remain prohibitively complex. Maximum a posteriori [21] and maximum likelihood [22] decoding can also enhance the receiver performance, though they sacrifice simplicity. In recent years, machine learning (ML) techniques have penetrated the area of nonlinear channel equalization with their ability to track intersymbol dependencies [23]–[25].

This work was supported by the Hellenic Foundation for Research and Innovation (H.F.R.I.) under the "2nd Call for H.F.R.I. Re-search Projects to support Faculty Members & Researchers" (Project Number: 2901) and the UK's EPSRC grants EP/S028854/1 and EP/P003990/1.

Stavros Deligiannidis, Kostas Sozos and Adonis Bogris are with the Department of Informatics and Computer Engineering, University of West Attica, Aghiou Spiridonos, Egaleo, 12243, Athens, Greece (e-mail: sdeligiannid@uniwa.gr, ksozoas@uniwa.gr, abogris@uniwa.gr).

K. R. H. Bottrill and P. Petropoulos are with the Optoelectronics Research Centre (ORC), University of Southampton, Southampton SO17 1BJ, United Kingdom. (e-mails: k.bottrill@soton.ac.uk; pp@orc.soton.ac.uk)

Charis Mesaritakis is with the Department of Information & Communication Systems Engineering, University of the Aegean, 2 Palama & Gorgyras St., 83200, Karlovassi Samos, Greece (e-mail: cmesar@aegean.gr).

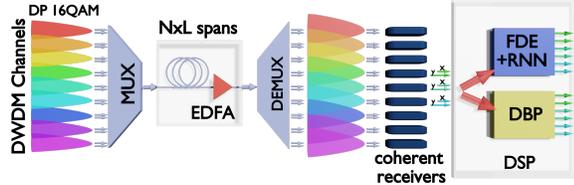

**Fig. 1:** The simulated 9-channel WDM transmission system containing a multi-channel coherent receiver. The DSP stage applies the two different equalizing techniques, DPB and RNN. In the case of RNN we first perform dispersion compensation using the frequency domain ideal equalizer (FDE)

Amongst other models, bidirectional Recurrent Neural Networks (bi-RNN) have proved efficient mitigation performance at moderate complexity [26], [27]. Furthermore, ML can also be used to improve the DBP method by introducing a learned digital backpropagation technique that relies on parameterizing the split-step Fourier method [28].

In this work, we propose the merge of multi-channel equalization with ML in a multi-channel bidirectional Vanilla RNN (bi-VRNN) equalizer. The equalizer takes advantage of multi-input multi-output (MIMO) processing and decoding of the WDM channels. In this way, it exploits useful information from adjacent channels for a better tracking of inter-channel dependencies. Although MIMO equalization based on deep learning has been proposed in the past [29], this previous work aimed at compensating mainly for FWM at orthogonal frequency division multiplexed signals, and demonstrated marginal complexity improvement with respect to DBP. Furthermore, the complexity increased proportionally with the parallel processing of adjacent channels. In the context of ML, Sidelnikov et al. [30] have also proposed an elegant multi-channel DBP solution with the use of CNNs which provides equivalent or even better than DBP performance and much less complexity. However, this work does not avoid the fact that the complexity scales with the number of spans. Our work is solely focused on the nonlinear mitigation of transmission impairments as linear effects can be easily handled by linear equalizers in the time or the frequency domain, which means that linear compensation is applied only once and not per each span as in the DBP case. Our proposition aims both at performance gains through the increase of useful information from adjacent WDM channels and complexity reduction when a multi-channel RNN equalizer is used instead of a single-channel one. RNN is more difficult to parallelize in FPGAs as shown in [31], however we believe it is worth using the bi-RNN proposed here as it requires a fairly low number of hidden units which can be served by the DSP [31]. Moreover, a recent work showed that using knowledge distillation one can resolve the parallelization problem of recurrent neural networks [32]. Here, we prove through extensive numerical simulations that the proposed approach outperforms both typical multi-channel equalization in the form of DBP and single-channel bi-RNN, offering BER improvement and/or Optical Signal to Noise Ratio (OSNR) gain. We experimentally verify this concept with a 180 km, 3-channel 11.25 Gbaud QPSK transmission system that mimics the nonlinear impairments of a long-haul transmission system by increasing the power to emulate inter-channel nonlinear effects accumulated over thousands of km.

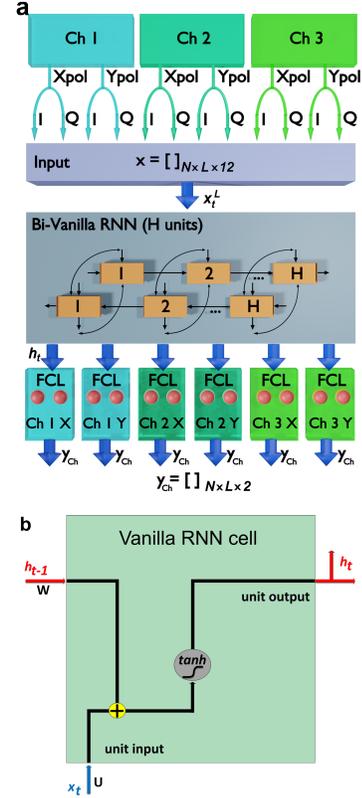

**Fig. 2:** The model structure of the Bidirectional Vanilla-RNN equalizer in the case of joint training and equalization of 3 WDM channels, where the bi-Vanilla-RNN layer consists of $H$ hidden units and b) the composition of a Vanilla-RNN hidden cell

Finally, we demonstrate competitive complexity savings, illustrating more than 550% reduction compared to the single-channel DBP in long-haul considered scenarios (>1000 km).

The paper is organized as follows: In section II, we describe the architecture and operation principles of the RNN and particularly introduce the multi-channel equalization approach. In section III, we present simulation results that demonstrate the multichannel RNN equalizer BER performance and show the advantages over the single channel RNN and DBP. In Section IV, we verify the improvement of multichannel RNN performance based on an experimental QPSK transmission system. Section V discusses further the aspects of computational complexity. Finally, we conclude the paper in Section VI.

## II. Model Description

### A. Multi-channel bi-VRNN

In this work, the simulations consider transmission in typical single-mode fibers (SMFs) in the C-band. Since optical fiber is a nonlinear dispersive channel, we also include simulations where dispersion is lower than that of typical SMFs. Although this case is not frequently met in commercially deployed links and could be only served by non-zero dispersion shifted fibers, it is scientifically important to identify how dispersion values affect multi-channel equalization performance, especially if one takes into account the prospect of expanding long-haul transmission systems to other bands, such as O-band. Performance is evaluated by means of BER which is calculated through error counting at the receiver. The schematic of our model transmission system depicted in Fig.1, corresponds to 1200 km transmission, simulated with the integration of Nonlinear Schrodinger equation (NLSE), considering a fiber with attenuation coefficient $α=0.2$ dB/km, second order dispersion coefficient $β_2=-20$ $ps^2$/km and Kerr nonlinear parameter $γ =1.3$ $W^{-1}km^{-1}$. Fiber propagation is modelled based on Manakov's equations using the split-step Fourier method [33]. We consider 9-channel WDM transmission on a 75-GHz wavelength grid and dual polarization 16-QAM modulation at 64 Gbaud. Lumped periodic amplification is modelled with a noise figure of 5 dB and a span length of 50 km. Prior to any post-processing or demodulation, we perform chromatic dispersion compensation with the use of an ideal frequency domain equalizer (FDE) in the case of RNN equalizer. Polarization demultiplexing and carrier synchronization are ideally carried out as well to focus solely on the nonlinear impairments. The simulations are conducted with pulse shaping incorporating root-raised-cosine (RRC) shaping, with a roll-off factor of 0.1 and matched-filtering at the receiver. Finally, one sample per symbol (sps) is sent to the RNN processing unit.

The layered diagram of the bi-VRNN equalizer is shown in Fig. 2a for three co-processed WDM channels. We use both polarization components of 3 (or more) adjacent channels as an input to our neural network in order to apply joint equalization of multiple WDM channels. In [34], we had used only the two polarization components of the channel of interest as the input to the neural network model instead, which provides less information to the equalization process. The input $x_t$ to the bi-VRNN is a $N×L×12$ vector representing the multi-channel 16QAM signal, where $N$ denotes the total number of input symbols, and $L = 2k + 1$ stands for the length of the input. The input $x$ has a total of 12 features; the I and Q for both polarizations of each one of the 3 channels. At time $t$, the input can be expressed as

$$x_t^L = [ x_{t-k}, …, x_{t-1}, x_t, x_{t+1}, …, x_{t+k} ] \quad (1)$$

denoting that $k$ preceding and $k$ succeeding symbols to the current symbol $x_t$ vector are used to track the inter-symbol dependencies. The length $L$ depends on the foreseen channel memory, which relates to the accumulated dispersion in the SMF and the bandwidth limitation of the transceiver. The input $x$ goes through the bi-VRNN with $H$ hidden units, which process the input in both the forward and backward directions. According to the composition of a Vanilla-RNN's hidden cell, fig 2b, each hidden output $h_t$ is given by:

$$h_t = \tanh(W \cdot h_{t-1} + U \cdot x_t) \quad (2)$$

where $x_t \in \mathbb{R}^L$ is the $L$-dimensional input vector at time $t$, $H$ is the number of hidden units, $U \in \mathbb{R}^{H×L}$ and $W \in \mathbb{R}^{H×H}$ representing the trainable weight matrices; $h_t$, $h_{t-1}$ are the hidden output and previous hidden output, respectively; and $tanh$ denotes the hyperbolic tangent activation function. The hidden units are connected to each other through a recurrent structure. The bi-VRNN layer consists of two $H×H$ weight matrices, one for the forward direction and one for the backward direction.

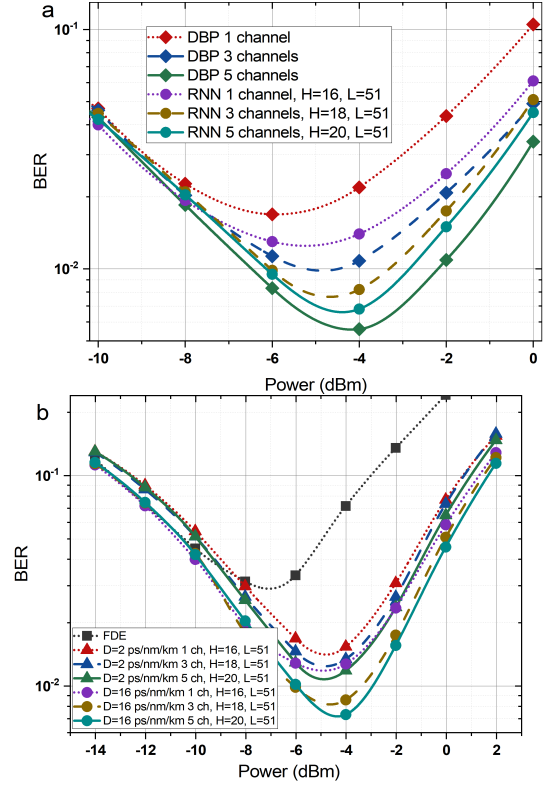

**Fig. 3:** BER as a function of launched optical power for single-channel and multi-channel processing, for RNN vs DBP in the C-band (a). In (b) RNN performance is evaluated at two dispersion regimes (b), respectively. Only the central's channel BER performance is plotted. Adjacent channels exhibit similar performance. <u>$H$ and $L$ the number of hidden units and the length of the word of symbols respectively</u>

The two directions are concatenated at the bi-VRNN layer's output $h_t$ with a combined weight matrix ($2H×2$). The same output of the bi-VRNN layer, $h_t$, is then sent in parallel to six (6) Fully Connected Layers (FCLs), which correspond to both polarizations of each one of the 3 channels jointly detected. Each FCL consists of two-state neurons, representing the $I$ and $Q$ of 16-QAM and its output $y$ is a $N×L×2$ vector. The FCL adopts the regression approach, which significantly reduces the implementation complexity without sacrificing BER performance, compared to the symbol-wise classification approach [34]. The bi-VRNN is trained using the *many-to-many* approach which produces simultaneously the same number of symbols $y$ as those at the input $x$ and is capable of extracting multiple symbols concurrently providing uniform BER for at least the 80% of the symbols aggregated in a word of symbols (41 over 51 in our case) [34]. The bi-VRNN equalizer was built, trained and evaluated in Keras with Tensorflow 2.10 GPU backend. We use Adam as an optimizer with a learning rate of $5×10^{-4}$. The mean square error (MSE) is chosen as the loss function for both training and validation. To avoid overfitting during training, we use early stopping when the validation loss does not decrease for 50 consecutive epochs. The weights of both bi-Vanilla-RNN and FCL are initialized using the Glorot uniform initializer [35]. During training, we consider 100 k words of symbols for training and 50 k words for validation while the $i$-th word is a shift of the ($i$-1)th word by one symbol. In this manner we increase the number of words that can be used to train the model. In the inference stage the shift is 40 symbols between successive words resulting in approximately 2.5 k words. Therefore, BER is calculated considering 2.5 k words with unknown data based on the evaluation of the 41 central symbols of each word. The training stage was executed with batches of 512 words of symbols. We note that the bi-VRNN model is adopted, because it offers similar performance when compared to bi-LSTM or bi-GRU [34], or CNN based networks at a low number of hidden units (<50) [36]. Due to complexity considerations, in this work we adhere to a maximum of 40 hidden units.

*B. Full-Field Multi-channel Digital Backpropagation*

As a well-established benchmark of the proposed RNN-based equalization system, we employ the multi-channel full-field DBP. The full-field DBP is considered impractical for real-life implementations as it requires a receiver with the bandwidth being capable of tracking the entire multi-channel spectrum [14]–[16]. However, this is the optimal benchmarking scenario for the simulation results and any other DBP variant would be inferior performance-wise. Also, several works in the past have approached full-field DBP with the use of spectrally sliced coherent receivers which can digitize the full optical field well beyond the electrical bandwidth limitations [37]. In our case, for adjusting the number of channels in the back-propagated field, we select the channels of interest through an optical filtering process that also rejects the unwanted out-of-band ASE noise. We then sample with 4 sps this 375 GHz signal, meaning that we employ an ideal ADC with 4x375 Gsa/sec sampling rate. The application of a symmetrical DBP with 20 steps-per-span follows. While 2 sps and 2 or 4 steps-per-span would be more realistic, we use these rather unusual parameters in order to compare bi-RNN approach with an almost ideal, channel-aware and greedy in terms of bandwidth and simulation steps equalizer. We optimize the nonlinear parameter in each DBP simulation in order to acquire the lowest possible BER. Following the DBP, the channels are separated with digital filters, downconverted in baseband and downsampled to 1 sps. Finally, matched filtering is applied to each one with symbol de-mapping and BER counting to assess the transmission performance of the measured sub-channel.

### III. SIMULATION RESULTS

*A. Bi-VRNN versus DBP results*

In Fig. 3, the equalization results presented in the form of BER values as a function of the launched optical power are depicted after having accomplished training of the RNN model under different conditions. We consider 9-channel dense WDM transmission with 75 GHz wavelength grid for a typical SMF in the C-band (second order dispersion parameter $\beta_2$=-20 ps$^2$/km (Fig 3.a) and a low-dispersion fiber ($\beta_2$=- 2 ps$^2$/km) over 1200 km of fiber (Fig 3.b). We compare the BER improvement offered by the multi-channel detection scheme for 3 and 5 neighboring channels versus the application of the DBP equalization technique using a corresponding number of channels. In Fig.3 only the BER of the central channel is depicted. Similar behavior is recorded for all co-processed

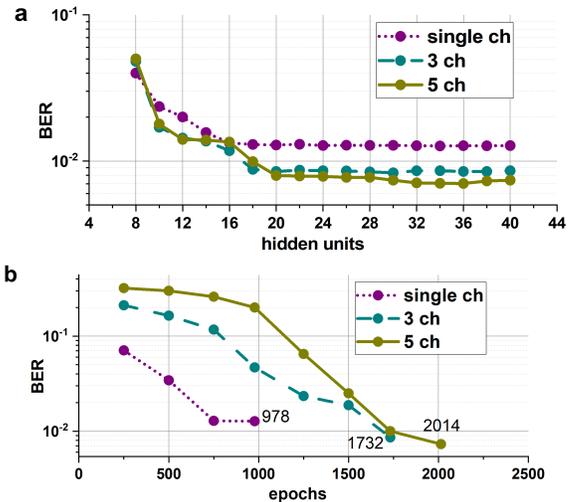

**Fig. 4:** BER as a function of a) bi-VRNN's hidden units and b) training epochs. 978, 1732 and 2014 training epochs are required for 1, 3 and 5 channels respectively

channels. Despite the fact that the low-dispersion channel memory is much smaller than that of the typical C-band regime, for the sake of equal comparison, the length of the input word ($L$) is set to 51 for both bands.

The utilization of multi-channel equalization through joint multi-channel training of the RNN model is observed to deal with the inter-channel nonlinear crosstalk caused by XPM and FWM effects, as bi-VRNN exhibits better performance than in the single-channel detection. The results show that for the C-band, the minimum BER for single-channel detection is $1.3 \times 10^{-2}$ @ -4 dBm, while multi-channel processing yields a minimum BER of $8.6 \times 10^{-3}$ @ -4 dBm and $7.3 \times 10^{-3}$ @ -3.5 dBm for 3 or 5 co-processed channels, respectively, as illustrated in Fig. 3a. Compared to DBP equalizers, which demonstrate improved performance with an increasing number of channels (minimum BER of $1.9 \times 10^{-2}$ @ -6 dBm, $1.1 \times 10^{-2}$ @ -6 dBm and $6.9 \times 10^{-3}$ @ -4 dBm, for 1, 3 and 5 channel respectively), the training and evaluation of 5 adjacent channels using bi-VRNN performs slightly better than the use of 3 channels in DBP which is operated almost ideally here in terms of the samples per symbol and transmission steps, as discussed in Section II. Therefore, the multi-channel bi-RNN equalizer achieves almost ideal equalization performance similar to that offered by a greedy multi-channel DBP implementation.

While the equalization performance shows a fair improvement in the high dispersion regime when multi-channel equalization is used, in the case of low-dispersion regime, the contribution of neighboring channels to the equalization performance is less beneficial as shown in fig. 3b. Thus, besides the significant impact of adjacent channels as the dispersion decreases, the coherence time of the channel also decreases resulting in inter-channel crosstalk with a high frequency content comparable to the symbol rate. In such scenarios, the RNN model is not able to learn from inter-channel interference and incorporate it into the equalization process. This finding is similar to what had been observed in [26] when the BER performance of bi-RNN was investigated at different dispersion regimes for single-channel equalization. It had been become evident that the nonlinear equalization performance of bi-RNN was superior as channel memory increased due to dispersion. Consequently, the burden of training the model with the neighboring channels in a low-dispersion environment compared to the C-band scenario, results in a marginal improvement in terms of BER. In specific, a minimum BER of $1.5 \times 10^{-2}$ @ -4 dBms, $1.3 \times 10^{-2}$ @ -4 dBm and $1.2 \times 10^{-2}$ @ -4 dBm, for 1, 3 and 5 channels respectively, was observed. The BER results occur after an optimization process which always takes into account keeping the number of hidden units (h.u.) below 25 in order to preserve complexity at moderate levels.

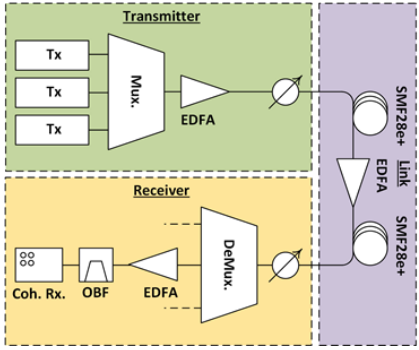

**Fig. 5:** The experimental transmission system

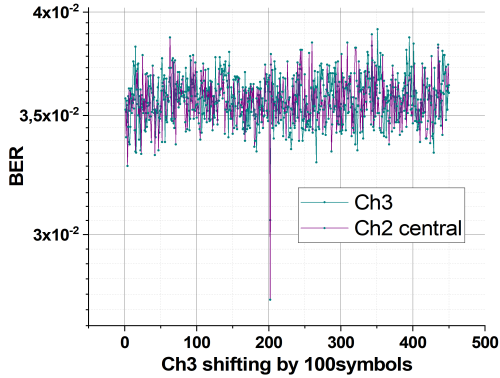

**Fig. 6:** Brute-force technique for temporally aligning the three independently captured (20 dBm launched power, -10 dBm receiver power). When temporal alignment was achieved, joint equalization improved BER from an average of $3.6 \times 10^{-2}$ to a sharp dip of $2.5 \times 10^{-2}$.

*B. Dependence of the RNN on hidden units and epochs*

In the optimization of the RNN equalization capacity, expressed by the number of its h.u. *H,* we validate the reasonable fact, that the gradual addition of h.u is necessary as the number of training channels increases. In fig. 4a it can be concluded that 16, 18 and 20 h.u are required for reaching the minimum BER performance for the single and multi-channel case of 3 or 5 channels respectively. Providing a larger network through increasing the number of h.u. does not improve the BER performance any further. Similarly, training the multi-channel network requires a larger number of epochs compared to the single-case scenario, fig. 4b. Using the early stopping mechanism, the optimal BER is reached at 978, 1732 and 2014 training epochs for 1, 3 and 5 channels. This can be explained by the fact that the additional channels increase the information at the input, and therefore a larger number of weights must be utilized to perform satisfactory equalization. It must be noted that the slight increase in h.u is not a severe shortcoming in terms of complexity as it will be shown in section V. Moreover, for static point to point WDM transmission systems where co-propagating channels remain stable over time, the substantial increase in training complexity is not a major issue, as training is not expected to take place very frequently in a real-life scenario [26].

IV. EXPERIMENTAL RESULTS

*A. Experimental Setup*

In this section we present experimental results which confirm the numerically derived findings. The experimental transmission system depicted in Fig. 5 consists of three independent transmitters, each composed of a ~25 kHz linewidth laser, followed by IQ-modulators that were themselves driven by arbitrary waveform generators (AWGs). The end result is three channels, CH1, CH2 and CH3, sitting at frequencies of 193.5 THz, 193.6 THz and 193.7 THz, respectively. Due to the memory limitation of the arbitrary waveform generator (DAC) and the use of the radix-2 FFT to perform pulse-shaping, each channel is delivering $2^{16}$ symbols of uncorrelated, random data using single-polarization, 22.5 Gbit/s QPSK signaling. We carefully selected the process for providing pseudorandom unrepeated sequences using rng('shuffle') and the very long period of $2^{19937}-1$ Mersenne Twister generator (as in simulations), so as to avoid bi-RNN to predict the next symbol of the pseudorandom sequence and overestimate the nonlinearity mitigation results. After signal generation, the three channels are co-polarized with polarization controllers before being multiplexed using an arrayed waveguide grating and amplified with an erbium doped fiber amplifier (EDFA). An attenuator allows for the control of the power launched into the transmission link. Transmission is carried out over a part of the UK's National Dark Fibre Facility (NDFF) and comprises two 91.9 km spans of field deployed SMF28e+ fiber on the route Southampton-Reading-Southampton. The signals are amplified with an EDFA mid-link such that the total power launched into the first and the second span of SMF28e+ will be the same. The receiver contains a variable optical attenuator to facilitate receiver characterization, a demultiplexing stage to select out the channel under measurement, an EDFA to optically pre-amplify the signal and an optical bandpass filter (OBF) in order to reject out-of-band noise. The signal is finally captured using a coherent receiver of typical construction, consisting of an optical hybrid followed by 4 balanced photoreceivers whose output is measured by a 40 GSa/s digital storage oscilloscope. The local oscillator and the carrier for each channel were sourced from the same original lasers. Once the waveforms had been captured (one at a time) for a range of powers launched into the link, they could be processed offline. It must be stressed out that although the experimental link is comprised from only two spans, we counterbalance by increasing the launched power in order to acquire a harsh non-linear environment and thus, mimic a longer transmission distance.

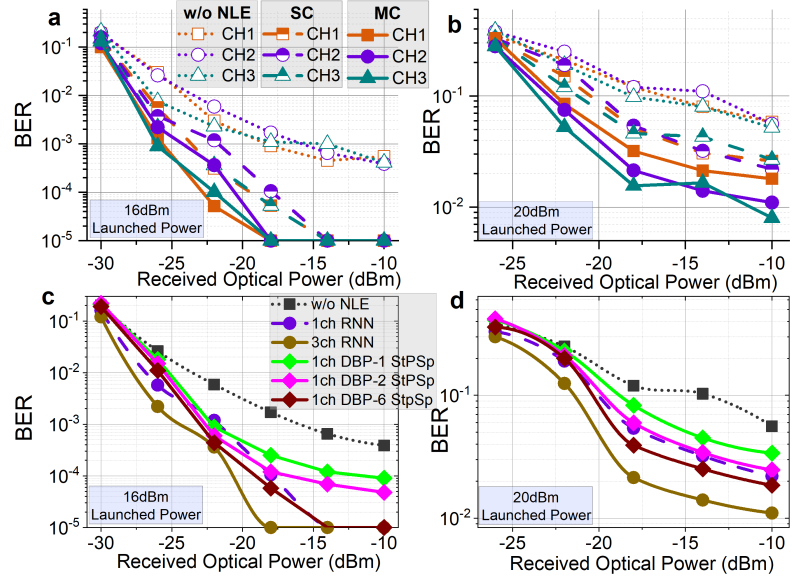

**Fig. 7:** BER as a function of received optical power for single- and multi-channel processing when the transmitted power per channel is 16 dBm (a), (c) and 20 dBm (b), (d) respectively. In (a) and (b) the performance of all transmitted and detected channels is depicted, whilst in (c) and (d) the performance for RNN, DBP or without any non-linear equalization (w/o NLE) considering the central channel is depicted. The number of hidden units $H$=16, 18 for SC and MC respectively and $L$=51 and the length of the word of symbols

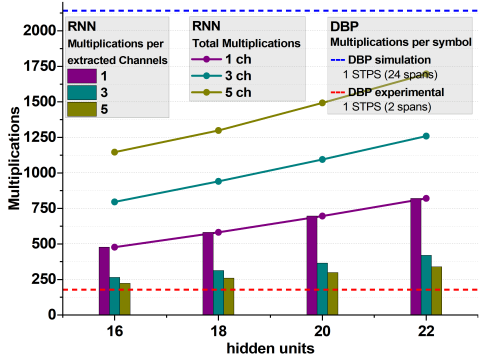

**Fig. 8:** Number of multiplications vs hidden units for single and multi-channel equalization.

*B. DSP Processing*

Prior to any nonlinear postprocessing or demodulation, we perform polarization alignment, bulk chromatic dispersion compensation with the use of a FDE and adaptive equalization using constant modulus algorithm (CMA) with 15 taps. We then perform carrier synchronization, clock extraction, and finally resampling. Eventually, one sample per symbol is sent to the bi-RNN processing unit. The only difference in the case of DBP post-processing is the removal of the FDE block, which instead is embedded in each step of the DBP process. The DBP is performed with 2 sps and 6 steps-per-span, before the CMA.

We use the 3 neighboring single-polarization channels as an input to our neural network in order to apply joint equalization. Its input $x$ is a $N \times L \times 6$ vector of QPSK symbols where $N$ denotes the total number of input symbols, and $L = 2k + 1$ stands for the length of the input, similarly to what we followed in the numerical analysis, with a proper modification from dual polarization 16-QAM to single-polarization QPSK. The same output of the bi-VRNN layer is then sent in parallel to 3 Fully Connected Layers (FCL), which correspond to each one of the 3 channels. Each FCL performs regression and consists of two neurons, representing the I and Q of QPSK. We considered 50,000 words of symbols for training, 5,000 words of symbols for validation, and 10,000 symbols for testing with unknown data.

The experimental process provided data for a wide range of input and received power values per channel. This data enables the direct derivation of BER performance for each channel when only linear equalization is performed or after the bi-RNN single channel equalization. Multi-channel equalization is not a trivial task to be carried out experimentally, since the data to be processed offline by the bi-RNN algorithm were collected in an asynchronous manner from the oscilloscope. In order to temporally align the three independently captured channels to the model, we had to use a time-consuming brute-force technique to identify the exact timing by performing joint equalization, initially for two out of three channels, for all possible combinations of their relative temporal position. For instance, as depicted in the example of fig. 6, by focusing on inter-channel effects between ch2-ch3 and by applying joint bi-VRNN equalization for different time shifts, we can identify the point where the BER is evidently improved as a result of a better temporal alignment of the two channels in terms of the inter-channel effects recognized by the RNN model. This process had to be repeated several times for fine-tuning (in fig. 6 the time-shift step is 100 symbols while at a second stage shifts of up to 10 symbols were carried out for precise synchronization) and for all the operating conditions of the optical power at the input and output. Through this brute-force method it was proven that there is a relative temporal position among all channels which results in a BER improvement for multi-channel processing (fig. 6), thus experimentally confirming the capability of the bi-RNN model to track inter-channel inter-dependencies. Indicatively, at a launch power of 20 dBm and a received power of -10 dBm, we observe a noticeable decrease in the BER from an average value of $3.6 \times 10^{-2}$ in temporally unrelated channels to $2.5 \times 10^{-2}$. It must be noted that in a practical system, the detection of co-propagating channels will be synchronous, therefore the data will be directly sent to bi-RNN for training and inference without the need of temporal alignment.

*C. Multi-channel versus single-channel bi-VRNN and DBP*

In Fig. 7, one can study the performance of bi-VRNN in single-channel and multi-channel equalization based on experimental data. The equalization performance exhibits BER improvement of up to an order of magnitude compared to single-channel processing. For instance, considering ch2 (central) at a launched power of 16 dBm (fig. 7a), the BER improves from $1 \times 10^{-3}$ to $4 \times 10^{-4}$. Even at the operating point of stronger nonlinear interaction (20 dBm launched power, fig. 7b), the joint equalizer can still provide a fair improvement in BER performance at all the received power values. In fig. 7a and 7b we notice that all 3 channels benefit almost equally from the joint processing. In the scenario of a launched power of 20 dBm and a received power of -10 dBm, we notice a reduction in the BER from $2.8 \times 10^{-2}$ to $1.8 \times 10^{-2}$ for Ch1 and from $2.7 \times 10^{-2}$ to $8 \times 10^{-3}$ for Ch3, respectively, upon applying single-channel and multi-channel equalization. In this harsh nonlinear environment caused by the high launched power nonlinear phase noise is very strong and the equalizer can not provide BER improvement of an order of magnitude. On the contrary, when launched power is 16 dBm, the improvement is very significant and the performance gain in terms of relative BER improvement is significant for all channels. Although the experimental conditions resemble that of a low dispersion system as only two spans are the case, the equalizer has a better performance than that evaluated in the simulations, as all co-propagating channels enter the equalizer and not only a portion of them (3 out of 9 in the simulation example).

The fact that the single-channel RNN approaches DBP with multiple steps-per-span (fig. 7c-d), indicates that it constitutes a near-optimum nonlinear equalizer of intra-channel effects. The further improvement that the 3-channel RNN offers, can therefore be attributed to the efficient XPM mitigation, rather than in residual SPM compensation. In fig. 7d, for the 20 dBm launched power the optical signal-to-noise ratio (OSNR) gain compared to single-channel bi-RNN is almost 7.5 dB @ BER=$2 \times 10^{-2}$, whilst for the 16 dBm launched power (fig. 7c), the OSNR gain is close to 2.5 dB @ BER=$10^{-3}$.

## V. COMPLEXITY ANALYSIS

In order to evaluate the computational complexity of multichannel equalization, we calculate the number of multiplications. The inference computational complexity, expressed as the number of multiplications following the many-to-many approach is expressed as [34]:

$$bi-VRNN_{mult} = 2(FH + H^2)L + 2HLy \quad (3)$$

where $F, y$ are the input features and output vectors respectively. The parameters $F$ and $y$ are equal to $2m$, since we have 2 inputs and outputs (I/Q components) for each channel, where *m* represents the number of jointly equalized channels (number of polarization modes multiplied by the number of WDM channels equalized). We use $H$=16, 18 and 20 h.u. for single and joint detection of 3 and 5 neighbouring channels. Based on the fact that at least 80% of neighbouring symbols exhibit a similar BER in the many-to-many approach [34] and since we have used a training word length of 51 symbols, we utilize the 41 central ones to calculate the number of multiplications in total. In fig. 8 we plot the total number of inference multiplications for the single and the multichannel cases versus the bi-VRNN's hidden units. One can see that although the number of multiplications increases with the addition of more input/output channels (features), eventually the number of multiplications per symbol decreases drastically. Dividing eq. 3 by the number of polarization modes multiplied by the number of WDM channels equalized (*m*) we calculate 478 multiplications per detected symbol (mps) for the single channel of 16 h.u, 313 mps for 3 extracted channels of 18 h.u. (34.38% complexity reduction compared to single channel bi-RNN), and 299 mps for 5 extracted channels of 20 h.u. (37.5% complexity reduction compared to single channel bi-RNN). The complexity reduction of multi-channel equalization over the single-channel counterpart in the scenarios we have considered can reach up to 58.7% in the case of 22 h.u with 821mps for the 1-channel relative to 339 mps for the 5-channel detection (fig. 8). The interpretation of the reduction of computational complexity in the case of multichannel RNN equalizer is found in (3) which reveals that most multiplications relate to the number of h.u. In fact, by taking advantage of the structure of the network and by adding a reasonable overhead (additional input and extracting channels), we manage to reduce the DSP processing per symbol whilst significantly improving the BER. Figures 3 and 4 demonstrate that the use of more than 3 channels contributes less to the improvement of the BER. Although it is out of the scope of this study, the hardware implementation of RNN equalizers in optical communication systems is a challenge that already concerns the research community [38]. We believe that the DSP implementation of a three-channel bi-RNN equalizer requiring a rather reasonable number of h.u. is feasible. This will be the subject of future work.

TABLE I
COMPLEXITY ANALYSIS

| Experiment (2 spans, QPSK single polarization) | Multiplications per symbol |
|---|---|
| Single-Channel DBP 1 step per span | 178 |
| Single-Channel DBP 2 steps per span | 357 |
| Single-Channel DBP 6 steps per span | 1070 |
| Single Channel bi-VRNN, 16 h.u | 877 (796 RNN+81 FDE) |
| Three Channel bi-VRNN, 18 h.u | 529 (447 RNN + 81 FDE) |

| Simulation (24 spans, 16-QAM pol-mux) | Multiplications per symbol |
|---|---|
| Single Channel DBP 1 step per span | 2141 |
| Single Channel DBP 2 steps per span | 4282 |
| Single Channel DBP 6 steps per span | 12845 |
| Single Channel bi-VRNN, 16 h.u | 558 (477 RNN + 81 FDE) |
| Three Channel bi-VRNN, 18 h.u | 394 (313 RNN + 81 FDE) |
| Five Channel bi-VRNN, 20 h.u | 380 (299 RNN + 81 FDE) |

Concerning the computational complexity of the RNN in the experimental setup, we calculate 796 mps for the single extracted channel with 16 h.u. and 448 mps for the 3 extracted channels with 18 h.u. which is equivalent to 43.75% complexity reduction compared to the single channel RNN.

For the DBP algorithm the complexity per bit can be evaluated according to [39]:

$$C_{DBP} = 4N_{StpSp}N_{spans}\left[\frac{N(log_2N+1)n_s}{(N-N_d+1)} + n_s\right] \quad (4)$$

where $N_{Span}$ is the number of spans, $N_{StpSp}$ is the number of steps per span, $N$ is the FFT size (depending on accumulated dispersion per span), $n_s$ are the sps and $N_d = n_s\tau_D/T$, where $\tau_D$ corresponds to the dispersive channel impulse response and $T$ is the symbol duration. We multiply the complexity by 4 since for DBP we use complex numbers and one complex multiplication is equal to four real multiplications. Considering $N$=256 and $N_d$=30, with 1, 2 or 6 steps-per-span for the 2 spans, we calculate 178, 357 and 1070 mps.

In order to achieve a fair comparison between the bi-RNN and the DBP approach, it is imperative to incorporate into the former the 81 mps necessitated by the FDE, according to:

$$C_{FDE} = 4\left[\frac{N(log_2N+1)n_s}{(N-N_d+1)}\right] \quad (5)$$

Taking into account FDE, the multi-channel bi-RNN approach with 18 h.u (requires 529 mps) would offer over 42,37% complexity reduction per channel with respect to a simplified DBP implementation (918mps), when 3 channels are incorporated, and 88,32% reduction compared to the greedy 20 step-per-span DBP (4528mps) considered in this paper. Table I clearly show that even for two spans, our biRNN processor is less complex than DBP considering 6 STPS (almost 200% more efficient) and its complexity advantage is even larger (~ 550%) when the number of spans increases to 24 which is the simulation case we have investigated in this paper even if we consider 1 STPS DBP. It is important to highlight that the computational complexity of the bi-RNN is not strictly related to the transmission length, whereas the DBP complexity is bound on the number of spans. Definitely, for long-haul transmission scenarios consisting of many spans, our bi-RNN approach will be by far less complex than DBP as shown in Table I. Additionally, both DBP and FDE equalizer require a minimum of 2 sps, in contrast to bi-RNN, which demonstrates satisfactory performance with only 1sps.

## VI. CONCLUSION

The paper investigates a bi-RNN based equalizer for joint training and equalization of adjacent channels of a WDM coherent system. We provided simulation results considering a polarization multiplexed 16-QAM transmitter operating at 64 Gbaud and experimental results with a WDM QPSK coherent system operating at 22.5 Gb/s. The simulation results revealed that multi-channel nonlinear equalization by means of bi-RNN is meaningful and can provide a clear performance advantage especially in high dispersion regimes. The theoretical findings were confirmed by an experiment incorporating 3-channel equalization. The proposed multi-channel equalization scheme exhibits BER improvement of up to an order of magnitude compared to a single channel detection RNN equalizer while reducing the computational complexity per symbol beyond 43%. The experimental and simulation results showed superiority in BER improvement compared to 3-channel full-field DBP and a significant complexity advantage which can be a complexity reduction in the order of 500% and beyond in long-haul transmission systems. We believe this multi-channel equalizer can constitute a practical solution for enabling reach extension in point-to-point digital coherent transmission systems, while requiring a moderate increase in receiver complexity. Next steps of our research will be directed to the implementation of a 3-channel bi-RNN equalizer in a field-programmable gate array platform.